\documentclass[twoside,twocolumn,9pt]{article}

\usepackage{extsizes}
\usepackage[left=1.5cm, right=1.5cm, top=1.785cm, bottom=2.0cm]{geometry}

\usepackage[utf8]{inputenc}
\usepackage{amsmath,amssymb}
\usepackage{graphicx}
\usepackage{hyperref,doi}
\usepackage[numbers,sort&compress]{natbib}

\usepackage{hyperref}
\usepackage{paralist}

\usepackage[version=3]{mhchem}

\def\wall{\mathrm{w}}

\def\fluid{\mathrm{fluid}}

\def\cst{\mathrm{cst}}

\begin{document}

\title{Mesoscopic simulations of anisotropic chemically-powered nanomotors}

\author{Pierre de Buyl%
\footnote{ORCID:\protect\href{http://orcid.org/0000-0002-6640-6463}{0000-0002-6640-6463}}
}
\date{Instituut voor Theoretische Fysica, KU Leuven B-3001, Belgium}

\maketitle

\begin{abstract}
Chemically powered self-propelled colloids generate a motor force by converting locally a
source of energy into directed motion, a process that has been explored both in experiments
and in computational models.
The use of active colloids as building blocks for nanotechnology opens the doors to
interesting applications, provided we understand the behaviour of these elementary
constituents.
We build a consistent mesoscopic simulation model for self-propelled colloids of complex
shape with the aim of resolving the coupling between their translational and rotational
motion.
Considering a passive L-shaped colloidal particle, we study its Brownian dynamics and locate
its center of hydrodynamics, the tracking point at which translation and rotation decouple.
The active L particle displays the same circling trajectories that have been found
experimentally, a result which we compare with the Brownian dynamics model.
We put forward the role of hydrodynamics by comparing our results with a fluid model in
which the particles' velocities are reset randomly. There, the trajectories only display
random orientations.
We obtain these original simulation results without any parametrization of the algorithm,
which makes it a useful method for the preliminary study of active colloids, prior to
experimental work.
\end{abstract}

\section{Introduction}

Nano- to micro-meter scaled self-moving colloids have become a commonplace experimental
device in soft-matter research groups, using a variety of driving methods including
catalytic chemical reactions~\cite{howse_et_al_prl_2007,ke_et_al_jpc_a_2010} or laser
illumination~\cite{kummel_circular_prl_2013,volpe_microswimmers_2011}.
The interest for these self-propelled particles, hereafter termed generically nanomotors as
in Ref.~\cite{kapral_perspective_jcp_2013}, stems both from their envisioned applications in
nanotechnology~\cite{ebbens_opinion_2016,kapral_perspective_jcp_2013,wang_nanomachines_2013}
and from the fundamental scientific insights offered by such
devices~\cite{howse_et_al_prl_2007,palacci_sedimentation_2010,thakur_kapral_active_media_jcp_2011}.
Given their small sizes, typical of colloidal systems, the trajectories of nanomotors
combine self-propulsion with the thermal fluctuations that are the signature of Brownian
motion.
Non-spherical colloids provide, even in the absence of self-propulsion, a
nontrivial realisation of Brownian motion with a possibly non-diagonal
diffusion matrix that couples the translational and rotational degrees of
freedom~\cite{brenner_arbitrary_shape_1965,brenner_arbitrary_shape_1967}.
The detailed experimental study of Brownian dynamics for model anisotropic colloids is
relatively recent. The colloids have been realised either by
photolithography~\cite{chakrabarty_boomerang_2013} or by assembling spherical
components~\cite{kraft_complex_shape_2013}.
The combination of self-propulsion and anisotropic colloids results in qualitatively new
types of trajectories:
circling patterns for Janus doublets~\cite{ebbens_et_al_pre_2010} and L-shaped
particles~\cite{kummel_circular_prl_2013}, and gravitaxis for L-shaped
particles~\cite{ten_hagen_gravitaxis_l_2014}.
The self-assembly of active rods, acting as building blocks for supra-colloidal anisotropic
active particles displays a similar phenomenology~\cite{davies_wykes_self-assembly_2016}.

In order to assess the role played by the shape of a nanomotor, one can resort to
numerical simulations to avoid costly trial and error in experiments.
Brownian dynamics simulation methods require the {\em a priori} knowledge of the mobility
tensor (or resistance matrix) of a particle and cannot serve, by themselves, for the study
of the shape.
It is possible to compute directly the resistance matrix using Stokes equation, for instance
using finite element methods~\cite{voss_hydro_2018}.
Here, we use instead coarse grained hydrodynamic simulations to resolve directly the
dynamics of complex-shaped colloids and of the self-generated flows.
Our choice extends naturally to other situations, such as many-motor dynamics in which the
resistance matrix also depends on the distance and relative orientation to other motors, and
integrates well with the mesoscopic chemical kinetics.

In this article, we study the L-shaped active colloid (or L motor, for short), explored
experimentally by Kümmel {\em et al}~\cite{kummel_circular_prl_2013}, using Molecular
Dynamics simulations.
We describe in section~\ref{modeling} the mesoscopic simulation
method~\cite{de_buyl_rmpcdmd_2017,ruckner_kapral_prl_2007}, in which the dynamics of both
the colloid and the surrounding fluid particles is resolved.
We present the bead model for the L shaped colloid in section~\ref{sec:L}.
In section~\ref{results}, we perform simulations of the equilibrium Brownian dynamics of the
L particle and identify the center of hydrodynamics for the colloid in the quasi-2D geometry
from the numerical data.
Second, we perform simulations of the L motor when the chemical activity induces
self-propulsion. We observe circling trajectories that result from the combined
translational force and the torque applied by the self-propelling force.
Third, we demonstrate the role played by self-generated hydrodynamic flows by considering an
alternative dynamical model for the fluid particles' dynamics, in which the conservation of
energy and momentum do not hold and no circling is observed.
Finally, we discuss our results in section~\ref{conclusions} and cast them in the
perspective of active matter research.

\section{Mesoscopic simulation method}
\label{modeling}

We consider a system with a single L-shaped colloid and a fluid of point particles.
The colloid consists in a rigid assembly of spherical beads of species $C$ or $N$. The
catalytic $C$ beads activate the conversion of fuel into product and the non-catalytic
$N$ beads are chemically inert.
The fluid consists of a large number of point particles of species $A$ and $B$, of mass $m$
for both species, with a number density per unit volume $\gamma$ and a mass density
$\rho=\gamma\ m$.

The colloid and the solvent in its vicinity evolve according to Molecular Dynamics, using
the velocity Verlet algorithm~\cite{allen_tildesley_1987,malevanets_kapral_mpcd_2000}.
To integrate the equations of motion for the colloid constructed as a bead assembly, we use
the quaternion-based velocity Verlet integrator described in
Ref.~\cite{rozmanov_rotational_2010}.
The dynamics of the solvent, in all regions where it is force-free, is given by the coarse
grained Multiparticle Collision Dynamics (MPCD) algorithm~\cite{malevanets_kapral_mpcd_1999}.
The fluid is partitioned every time interval $\tau$ in a lattice of cubic cells of side $a$
in which the velocities are exchanged according to an energy and momentum conserving rule.
We use the truncated Lennard-Jones potential to couple the colloids and the solvent
particles:
\begin{equation}
\label{VLJ}
V_{\kappa,\alpha}(r) =
4 \epsilon_{\kappa,\alpha} \left( \left(\frac{\sigma}{r}\right)^{12} - \left(\frac{\sigma}{r}\right)^{6} + \frac{1}{4} \right)
\end{equation}
for $r \le \sigma\times 2^{1/6}$. $\kappa$ denotes the colloid species, which can be either
``C'' for catalytic colloids or ``N'' for noncatalytic colloids, and $\alpha$ the solvent
species.
In the present work, we keep the scale parameter $\sigma$ equal for all species pairs and
set $\epsilon_{\kappa,\sigma}$ to non-equal values to obtain different relative repulsion
strengths at the surface of the colloid.
We use a quasi-2D geometry as the corresponding experiment was confined between two glass
plates.
The coordinates x and y are periodic and the coordinate z is bounded by walls at $z=0$ and
$z=L_z$.
For the fluid particles, the walls in the z direction perform bounce-back
collisions~\cite{allahyarov_gompper_mpcd_flows_2002,whitmer_luitjen_2010} and include ghost
particles during the collision step~\cite{lamura_mpcd_epl_2001}.
For the colloid, we apply a confinement potential in the z direction as
\begin{equation}
V(z) = \epsilon_{\wall} \left\{\frac{3\sqrt{3}}{2} \left( \left(\frac{\sigma_{\wall}}{z-z^\ast_\wall}\right)^{9} - \left(\frac{\sigma_{\wall}}{z-z^\ast_\wall}\right)^{3}\right) + 1 \right\} ~,
\end{equation}
where $\epsilon_{\wall}$ is the strength of the potential, $\sigma_\wall$ its scale and
$z^\ast_\wall$ is a shift of the potential's origin.

We use the units system consistently defined by $m=1$ for the mass,
$\epsilon_{N,A}=\epsilon_{C,A}= 1$ for the energy and $a=1$ for the length. The unit of time
is $a\sqrt{m/\epsilon}$.

To achieve a chemical reaction that is localised on the active part of the colloid, the
beads of species ``C'' trigger the chemical reaction
\begin{equation}
\label{catalytic}
\cee{A + C ->[p] B + C}
\end{equation}
in the fluid~\cite{ruckner_kapral_prl_2007}. The reaction is carried out with probability
$p$ when the triggered solvent particle exits of the interaction range of the colloid to
avoid discontinuities in the trajectory.
We keep the simulation out of chemical equilibrium with a bulk reaction, using the reactive
MPCD (RMPCD) algorithm~\cite{rohlf_et_al_rmpcd_2008,thakur_kapral_active_media_jcp_2011}
\begin{equation}
\label{bulk}
\cee{B ->[k_2] A} ~.
\end{equation}
The bulk reaction is carried cell-wise and, similarly to what is done for the catalytic
reaction~\eqref{catalytic}, cells for which one or more fluid particles are interacting with
colloids are excluded from the reaction process.

We list the simulation parameters for the fluid in table~\ref{fluid-params}, along with the
theoretical values for the transport coefficients of the fluid.
\begin{table}[ht]
\centering
\begin{tabular}{l l}
  Parameter & Value\\
  \hline
  \hline
  $k_B T$ & 1\\
  $\rho$ & 10\\
  $\tau$ & 0.1\\
  $\gamma$ & 10\\
  MPCD collision angle & $\pi/2$\\
  \hline
  \\
  Transport coefficients\\
  \hline
  \hline
  Viscosity $\eta$ & 5.43\\
  $D_\fluid$ & 0.117\\
  Sc & 4.7\\
  \hline
\end{tabular}
\caption{Simulation parameters for the fluid. The viscosity and self-diffusion coefficients
  are computed using formulas (52) and (43) of Ref.~\cite{kapral_adv_chem_phys_2008}.}
\label{fluid-params}
\end{table}

In order to assess the role of hydrodynamics, we also perform simulations in which we
replace the MPCD collision step by a random sampling
(RS)~\cite{belushkin_hydrodynamics_2012} of the fluid particles' velocities.
At fixed time intervals $\tau$, instead of colliding the particles cell-wise, we draw their
velocities from a thermal distribution.
With the RS method, we still follow the fluid particles' positions as they are necessary to
generate chemical concentration gradients in the system.
Using this alternative dynamical evolution for the fluid, we cannot match the dynamical
properties of the colloid to the ones of the MPCD
simulation~\cite{colberg_kapral_manybody_2017}.
Instead, we use it to compare qualitatively the colloid's dynamics with and without
hydrodynamics.

We perform all simulations with the open-source package
RMPCDMD~\cite{de_buyl_rmpcdmd_2017,rmpcdmd_1.0,rmpcdmd_web} and provide all the parameter
and analysis files in appendix~\ref{repro} and in the supplementary
information~\cite{complex_nanomotors_2018}.

\section{Bead model for the L particle}
\label{sec:L}

The simulation strategy outlined in section~\ref{modeling} was first demonstrated for dimer
nanomotors~\cite{ruckner_kapral_prl_2007}, in which only two beads make up the colloid. It
has also been used since for bead models, such as for the Janus
particle~\cite{de_buyl_kapral_nanoscale_2013}.
Chemically induced self-propulsion is well studied in simulations and has been analysed in
great detail~\cite{reigh_kapral_soft_matter_2015,reigh_janus_2016}, which makes it useful as
a reference.

Here, we design a L particle whose foot is catalytically coated, following the geometry of
Ref.~\cite{kummel_circular_prl_2013}.
In this latter work, however, the propulsion arises from the laser heating-induced demixion
of the solvent at the Au-coated side of the colloid, a self-propulsion mechanism pioneered
in Volpe {\em et al}~\cite{volpe_microswimmers_2011}.
In demixion-based propulsion, the local phase change in the fluid produces hydrodynamic
flows whose reciprocal effect enable the motion of the colloid.
Despite the difference in the type of motor, the hydrodynamic features of the present
self-diffusiophoretic mechanism fall in the same category of phoretic
motion~\cite{brady_jfm_2011}.
In both situations, the motor and the surrounding fluid are force-free and it is the
generation of flows that is responsible for self-propulsion.

\begin{figure}[ht!]
\centering
\includegraphics[width=.7\linewidth]{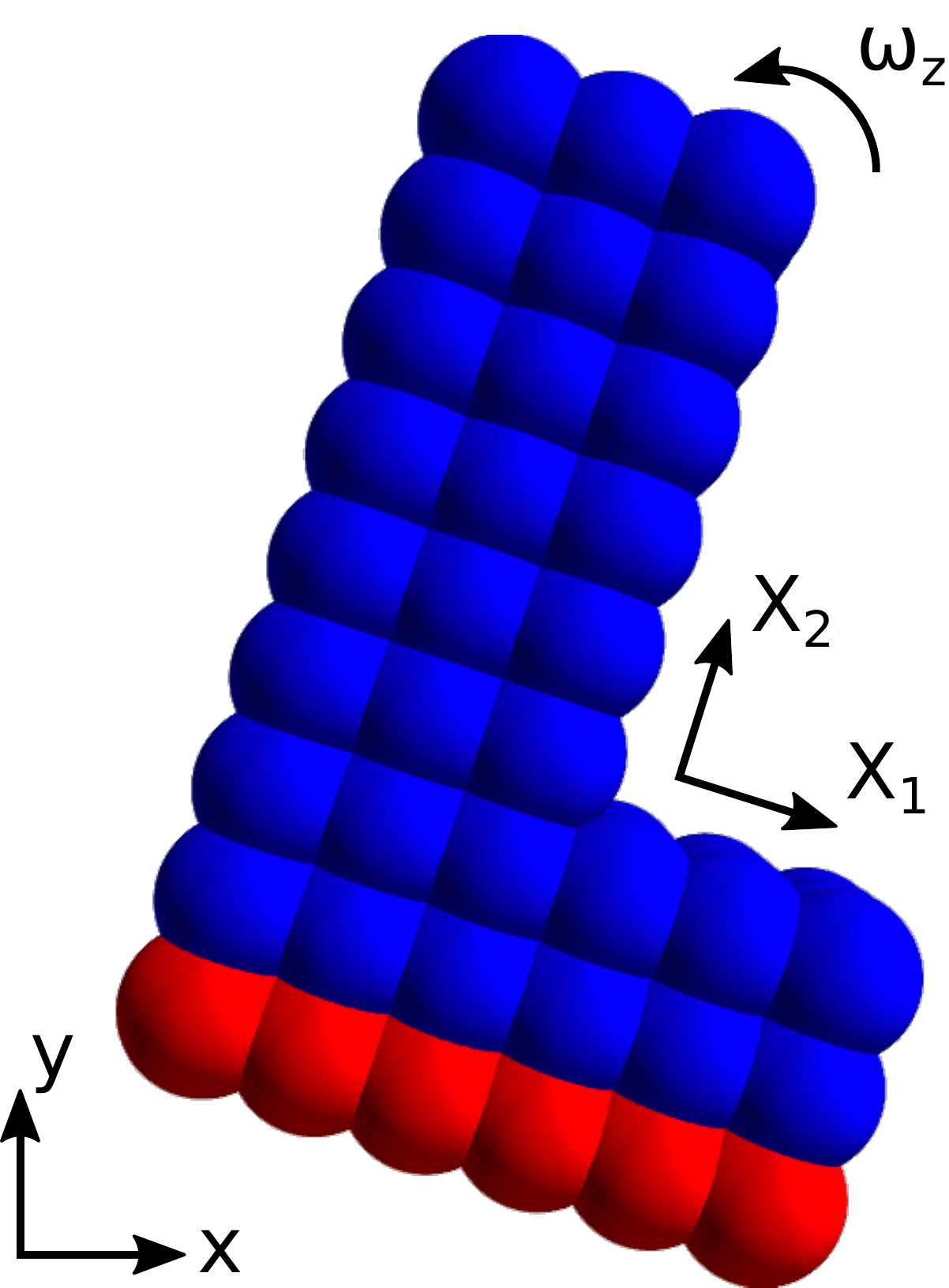}
\caption{The bead assembly for the L-shaped colloid. x and y denote the lab-frame
  coordinates, coinciding with the principal axes of the body for $t=0$. $X_1$ and $X_2$
  denote the unit vectors along the short and long arm of the L, respectively, and
  $\omega_z$ is the in-plane rotational velocity. We identify the third coordinate as the
  total angular displacement in the x-y plane: $X_3(t) \equiv \theta(t) = \int_0^t dt' \omega_z(t')$.
  There are three layers in the z direction, totalling 108 beads.}
\label{L}
\end{figure}

We assemble beads in a L-shaped colloid, formed by an array of equally spaced beads and
represented in Fig.~\ref{L}.
The figure shows the particle with its principal axes aligned with the laboratory frame
coordinates (the x-y coordinates of the figure).
As in Ref.~\cite{kummel_circular_prl_2013}, we also define a coordinate system aligned with
the arms of the L. We illustrate those coordinates in Fig.~\ref{L}.

\section{Results}
\label{results}

\begin{table}[ht]
\centering
\begin{tabular}{l l}
  Parameter & Value\\
  \hline
  \hline
  $L_x, L_y, L_z$ & 50, 50, 13\\
  Number of solvent particles & 325000\\
  Number of beads & 108\\
  MD time step & 0.01\\
  Simulation length & $10^6 \tau$\\
  Lennard-Jones $\sigma$ & 1.5\\
  $\epsilon_{\kappa B}$ & 5\\
  $k_2$ & 0.01\\
  $\epsilon_\wall$ & 1\\
  $\sigma_\wall$ & 1.5\\
  $y^\ast_\wall$ & 3.5\\
  Number of equilibrium simulations & 20\\
  Number of self-propelled simulations & 20\\
  \hline
\end{tabular}
\caption{Simulation parameters for the L particles. See table~\ref{fluid-params} for the
  parameters of the MPCD fluid.}
\label{l-params}
\end{table}

\subsection{Brownian dynamics of the L particle}
\label{brownian}

\begin{figure*}[h!]
\centering
\includegraphics[width=.9\linewidth]{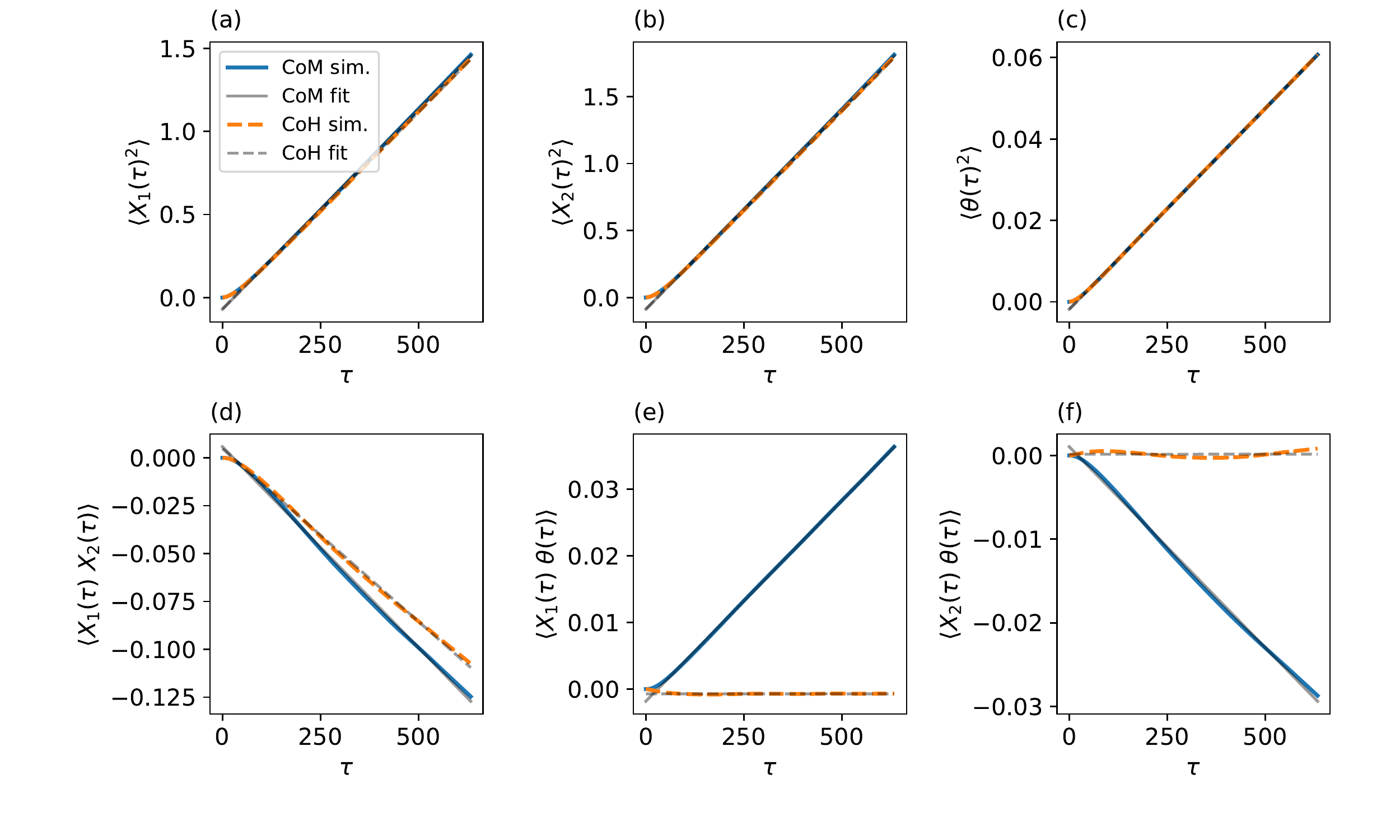}
\caption{The cross-displacements $C_{ij}(\tau)$ [Eq.~\eqref{eq:cij}] for the L particle in
  equilibrium. The thick lines show the simulation data when tracking the CoM (full line)
  and the CoH (dashed line). The shaded lines are the corresponding linear fits. When
  tracking the CoH, the translation-rotation coupling disappears [panels (e) and (f)].}
\label{L-cross-disp}
\end{figure*}

We perform equilibrium simulations of the L particle by setting the reaction probability
$p=0$. We list all simulation parameters in tables \ref{fluid-params} and \ref{l-params}.

Prior to showing the simulation results for the diffusion of the L particle, we introduce an
overdamped Langevin dynamical model for Brownian motion in two dimensions. The model is
equivalent to the one found in Ref.~\cite{kummel_circular_prl_2013} but we have chosen to
write it in the body-frame coordinates so that we can write the diffusion as a constant
matrix, following Ref.~\cite{chakrabarty_brownian2d_2014}.
The evolution of the positions is given by
\begin{equation}
\label{l-model}
\left(
\begin{array}{l}
  \dot X_1\\ \dot X_2\\ \dot X_3
\end{array}
\right)
= \sqrt{2 D^L} \zeta + \beta D^L F ~,
\end{equation}
where $X_1$ and $X_2$ are the coordinates shown in Fig.~\ref{L}, $X_3$ is the angle of the
particle in the x-y plane, $D^L$ is the diffusion matrix, $\zeta$ is a vector white noise
process and $F$ is an external force that models, when non-zero, the effect of
self-propulsion~\cite{kummel_circular_prl_2013,ten_hagen_force_torque_2015}.
We will use the model in Eq.~\eqref{l-model} for three purposes: define the measurements of
the diffusion matrix $D^L$, locate the center of hydrodynamics of the colloid, and estimate
the radius of the circles in the self-propelled regime.
We use the cross-displacement $C_{ij}(\tau)$, as defined in
Ref.~\cite{kraft_complex_shape_2013}:
\begin{equation}
\label{eq:cij}
C_{ij}(\tau) = \langle \left(X_i(\tau)-X_i(0)\right) \left(X_j(\tau)-X_j(0)\right) \rangle ~.
\end{equation}

The diagonal elements of $C_{ij}(\tau)$ are the usual mean-squared displacements that
represent diffusion. The coupling between the degrees of freedom results in non-zero {\em
  off-diagonal} elements in $C_{ij}(\tau)$.
From a fit of the linear-in-time evolution of the cross-displacement, we identify the
diffusion matrix
\begin{equation}
C_{ij}(\tau) \propto 2 D^L_{ij} \tau
\end{equation}
that quantifies the dynamics of the L particle
\cite{kraft_complex_shape_2013,chakrabarty_brownian2d_2014,chakrabarty_boomerang_2013}.

\begin{figure}[h!]
\centering
\includegraphics[width=.7\linewidth]{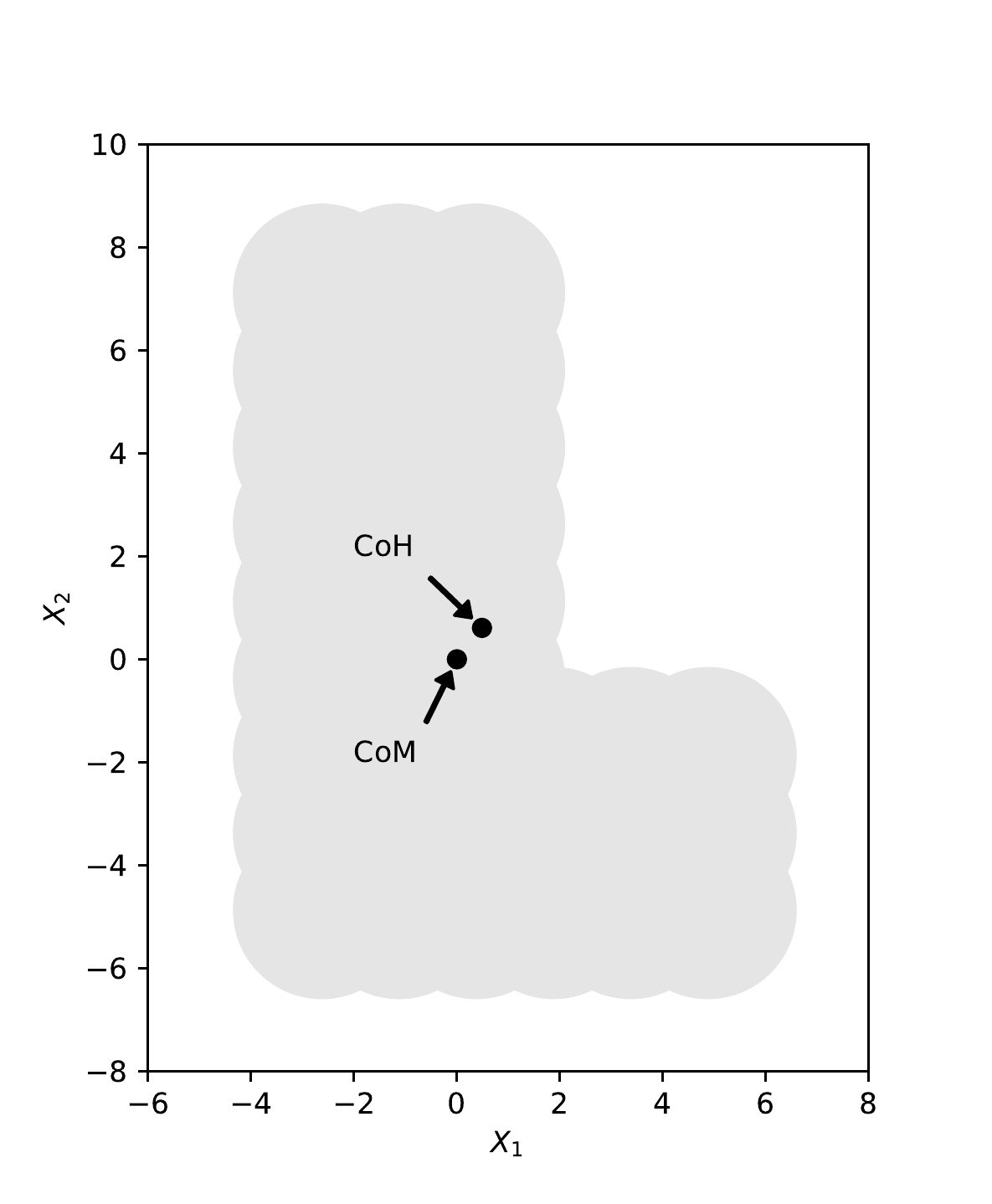}
\caption{Flat representation of the bead model with the center of mass (CoM) and the center
  of hydrodynamics (CoH).}
\label{L-centers}
\end{figure}

We show in Fig.~\ref{L-cross-disp} (full lines) the average cross-displacements of the L
particle in equilibrium, where we use the center of mass (CoM) as the tracking point of the
colloid. We obtain the following diffusion matrix by linear fits of the cross-displacement data:
\begin{equation}
\label{D-matrix}
D^{L@CoM} = \left(
\begin{array}{c c c}
12.04 & -1.05 & 0.30 \cr
-1.05 & 14.93 & -0.24 \cr
0.30 & -0.24 & 0.49 \cr
\end{array}\right) \times 10^{-4} ~.
\end{equation}
All degrees of freedom appear to be coupled. Apart for screw-type shapes, it is possible to
identify the Center of Hydrodynamics of the colloid, where the coupling between translation
and rotation disappear.
Chakrabarty {\em et al}~\cite{chakrabarty_brownian2d_2014} provide a method to find the CoH
for 2D Brownian colloids, which we apply here.
The relevant locations on the colloid are the center of mass (CoM), whose trajectory is
resolved in the simulation, and the center of hydrodynamics (CoH). Both locations are shown
in Fig.~\ref{L-centers}. By definition, at the CoH the coupling between translation and
rotation is zero in two dimensional
systems~\cite{happel_brenner_1991,chakrabarty_brownian2d_2014}.
For two-dimensional Brownian motion, Chakrabarty {\em et
  al}~\cite{chakrabarty_brownian2d_2014} have used an overdamped Langevin model to compute
the diffusion matrix at points other than the CoH (such as the experimental tracking point)
and also the shift necessary to locate the CoH.
In the reference frame of Fig.~\ref{L}, the relation between the two points is given by
\begin{equation}
\left\{\begin{array}{l l}
X_{1,CoH} &= X_{1,CoM}  + d_1\\
X_{2,CoH} &= X_{2,CoM} + d_2
\end{array}\right. ~,
\end{equation}
where $d_1 = -D^L_{2,\theta} / D^L_{\theta}$ and $d_2 = D^L_{1,\theta}/D^L_\theta$.

We show in Fig.~\ref{L-cross-disp} (dashed lines) the cross-displacements of the L particle,
using the center of hydrodynamics (CoH) as the tracking point. In comparison to the CoM
results, the angular displacement is now uncorrelated to the translational displacements.
Fitting the results of Fig.~\ref{L-cross-disp}, we obtain the diffusion matrix
$D^{L,CoH}$ at the CoH:
\begin{equation}
\label{D-coh}
D^{L@CoH} = \left(
\begin{array}{c c c}
11.86 & -0.90 & -0.00 \cr
-0.90 & 14.81 & 0.00 \cr
-0.00 & 0.00 & 0.49 \cr
\end{array}\right) \times 10^{-4} ~.
\end{equation}
To obtain the forces and torque exerted on the L particle by a self-propelling force
$F=\left(0, F, l~F\right)$, we need the lever length $l$ given by the distance between the
center of the catalytically coated foot of the colloid and the CoH. Considering the location
of the CoH $\left(X_{1,CoM} + d_1, X_{2,CoM} + d_2\right)$, the distance along axis $X_1$ to
the center of the foot of the L, where the force is applied, is $l=0.638$.
The CoH remains to the left of the colloid, in comparison to the center of the foot. We thus
expect the torque to generate anticlockwise rotation.

\subsection{Self-propulsion of the L-shaped particle}

\begin{figure}[ht!]
\centering
\includegraphics[width=\linewidth]{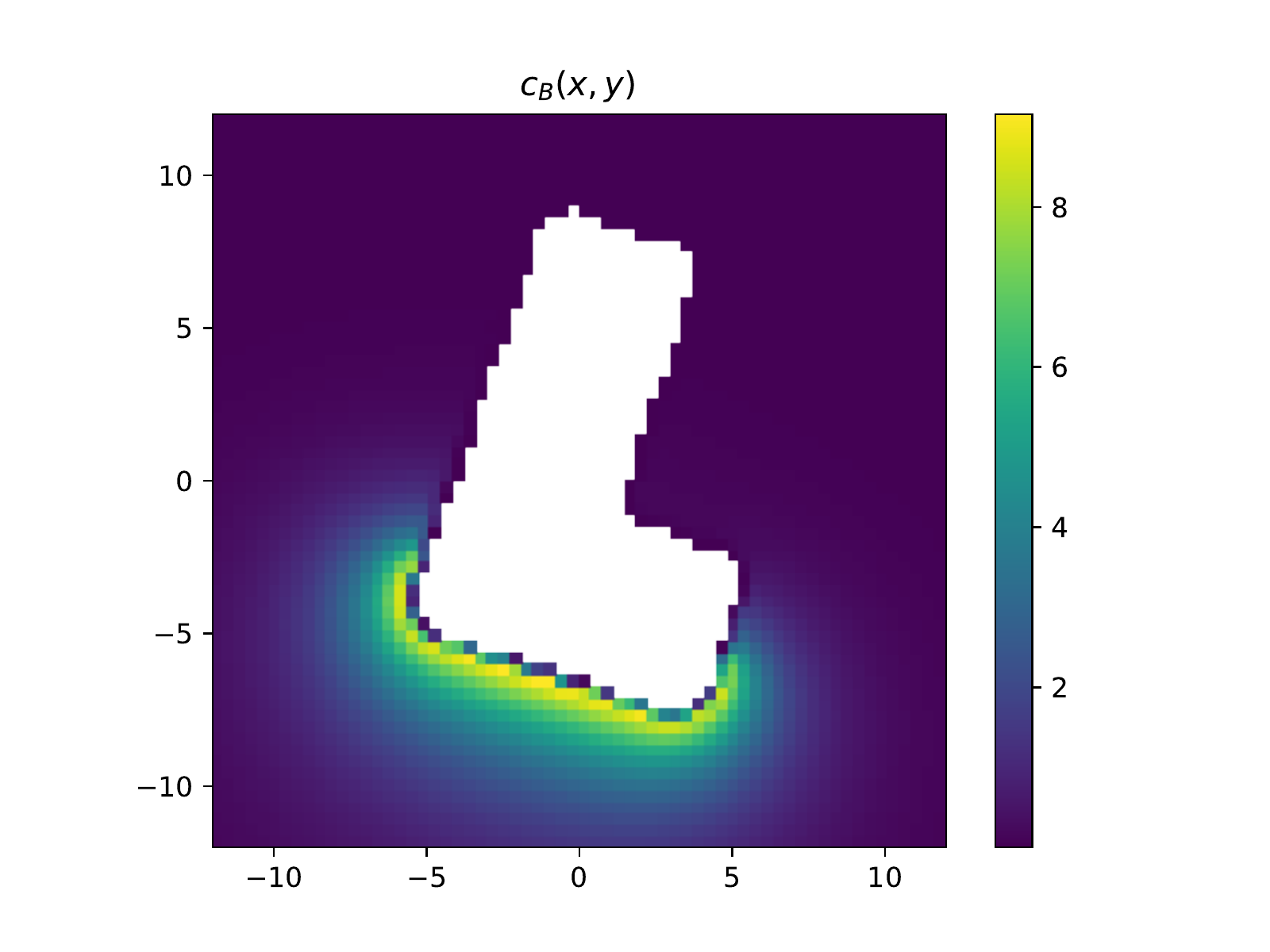}
\caption{The concentration of solvent species B around the L particle for a single
  self-propelled simulation, averaged over time. The solvent particles are sampled in the
  mid-plane of the L particles in a layer of thickness 1. The blank region is the inner
  volume of the colloid.}
\label{l-cb}
\includegraphics[width=\linewidth]{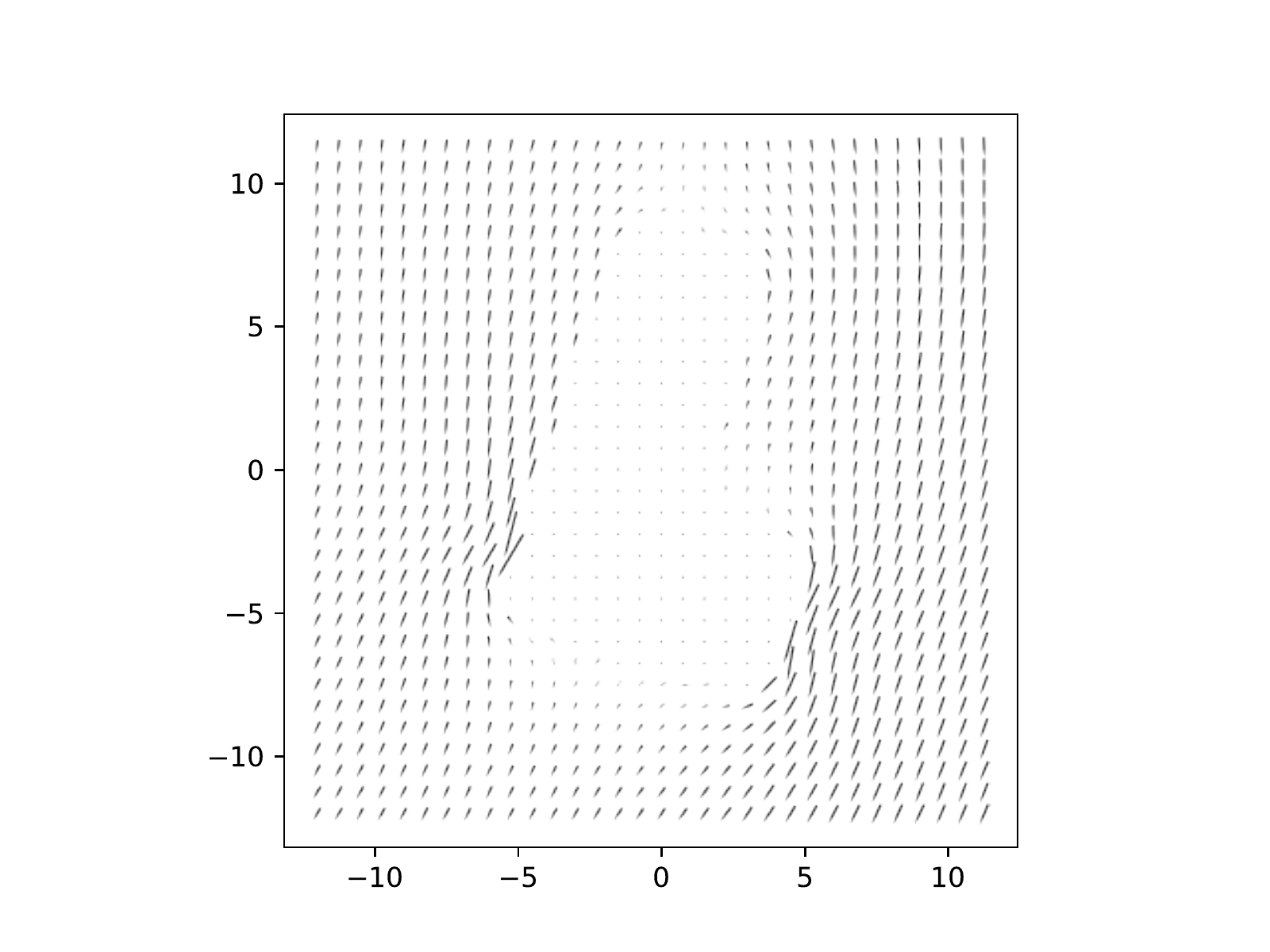}
\caption{Velocity field of the fluid around the L particle for a single self-propelled
  simulation, averaged over time. The sampling is performed relative to the colloid's center
  of mass velocity.}
\label{l-v}
\end{figure}

By enabling the catalytic reaction~\eqref{catalytic} (i.e. setting the reaction probability
to $p=1$), the fuel is converted on the active part of the L particle. We show the
concentration field $c_B$ of the B solvent species in Fig.~\ref{l-cb} and the velocity
profile of the fluid in Fig.~\ref{l-v}.
$c_B$ is maximal close to the catalytic surface of the L and decreases rapidly due to the
bulk reaction~\eqref{bulk}.

\begin{figure}[p]
\centering
\includegraphics[width=.75\linewidth]{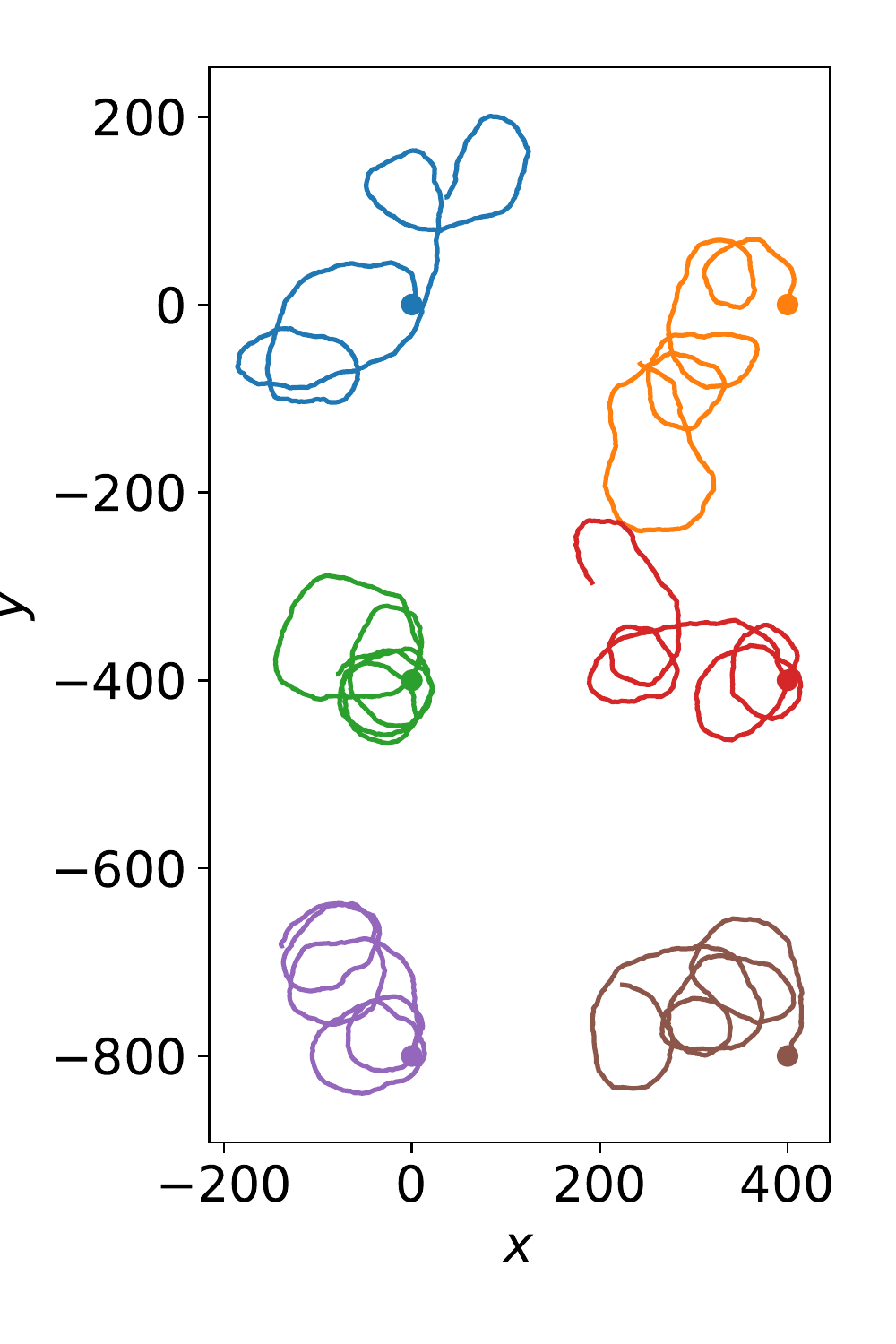}
\caption{Six example trajectories for the self-propelled L particle from independent
  simulations. The large dot for each trajectory indicates the starting point. The
  trajectories are shifted for display purposes.
  Circling trajectories appear for all simulation runs and are combined with thermal
  fluctuations.}
\label{l-sp-six}
\includegraphics[width=.75\linewidth]{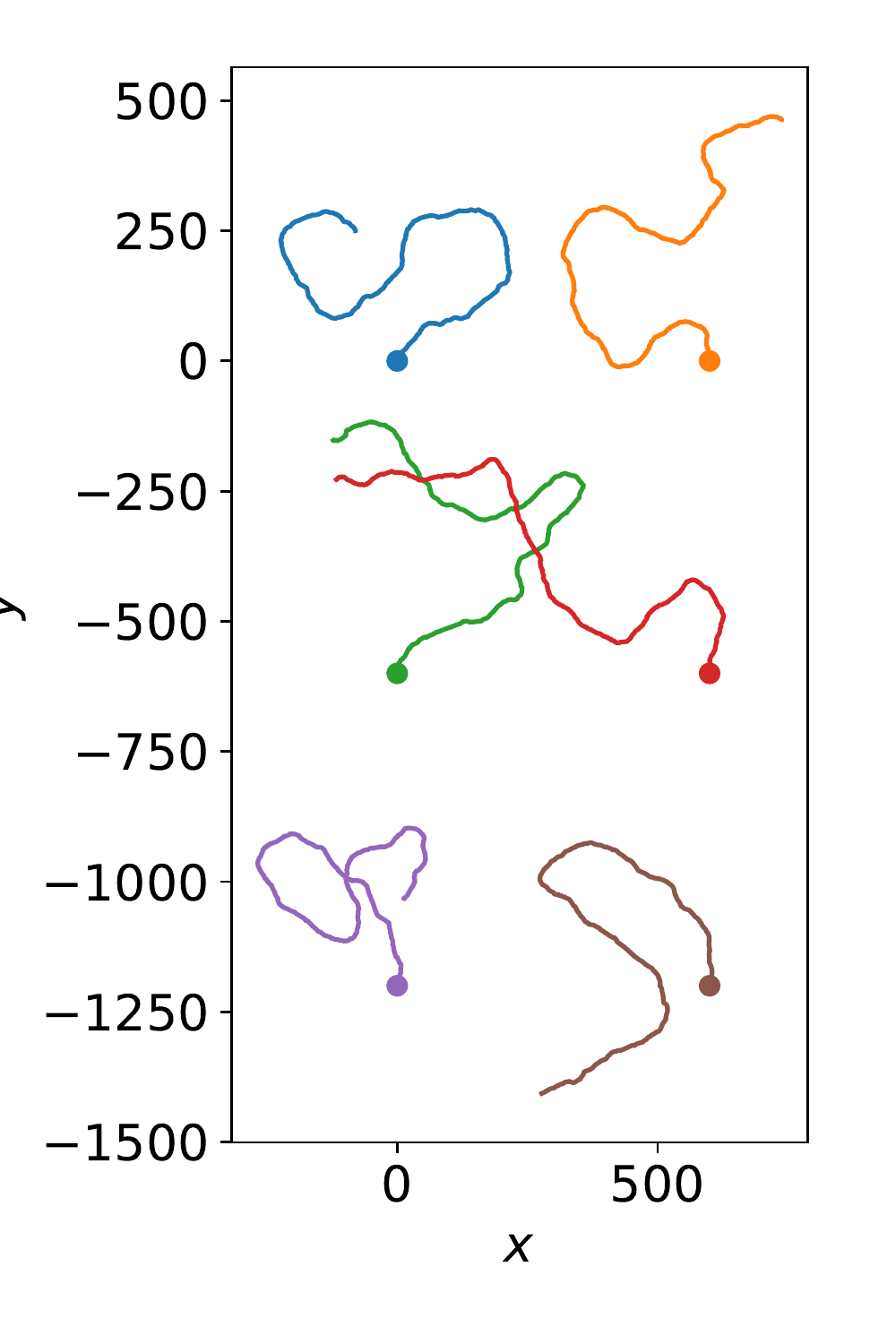}
\caption{Same as Fig.~\ref{l-sp-six} when hydrodynamics is turned off (RS fluid). There is
  no systematic orientation, at variance with the hydrodynamic (MPCD) simulations.}
\label{l-sp-six-noh}
\end{figure}

The L-particles, under nonequilibrium condition, display a self-propelling behaviour in
combination with thermal fluctuations.
We show example trajectories in Fig.~\ref{l-sp-six}, in which we notice that the motion
features a curved path. This is at variance with the Brownian motion of isotropic particles.

We highlight the systematic character of the rotation in Fig.~\ref{l-sp-six} by displaying
six of the trajectories for the self-propelled L particle.
Across all simulations, we compute the radius of the trajectories as the ratio of the
translational velocity to the angular velocity
$\omega_y$~\cite{kummel_circular_prl_2013}. The mean radius is $R \approx 55 \pm 7$, with
the error taken as the standard deviation over the 20 simulations.
We compare this value to the average velocity resulting from a constant force
\begin{equation}
F_\cst = \left(
\begin{array}{l}
  0\\
  F\\
  -l_L F
\end{array}
\right)
\end{equation}
on the motor in Eq.~\eqref{l-model}. The presence of the third element in $F_\cst$
originates in the torque due to the misalignment of the applied force on the foot of the L
with the center of hydrodynamics of the particle~\cite{kummel_circular_prl_2013}, with a
lever length $l$.
The average velocities are given by
\begin{equation}
\left(
\begin{array}{l}
  \dot X_1\\ \dot X_2\\ \dot X_3
\end{array}
\right)
=
\beta F D^{L@CoH} F_\cst
\end{equation}
and the radius is
\begin{equation}
R = \frac{\sqrt{\dot X_1^2 + \dot X_2^2}}{\dot X_3} \approx 47
\end{equation}
for a lever length $l=0.638$ (see section ~\ref{brownian}).
The diffusion matrix $D^{L@CoH}$ and the position of the CoH thus provide a reasonable
estimate of the rotation radius of the L nanomotor. The overdamped Langevin model does not
capture the hydrodynamic tails, which may limit its accuracy.

\subsection{The role of hydrodynamics in self-propulsion}

We present here simulations of the L nanomotor in which the fluid particles' dynamical
evolution follows the random sampling (RS) method.
We use equal parameters for the number density $\gamma=10$, the temperature $k_B T=1$ and
for the solvent-colloid interaction and the wall-colloid interaction (see
table~\ref{l-params}).
We resample the fluid's velocities every interval $\tau = 2$.

The simulations in Fig.~\ref{l-sp-six-noh} show that the motors move in spite of the fluid
not resolving hydrodynamic flows.
There is actually another possible self-propulsion mechanism that occurs when a
concentration gradient is combined with non-equal surface interaction, as it is the case
here.
This density-gradient induced self-propulsion, already investigated in
Ref.~\cite{ruckner_kapral_prl_2007}, usually contributes in addition to
self-diffusiophoresis but it is here the only source of active motion.

In the RS simulations, however, the friction arises from randomly moving solvent particles
that act as a rather expensive thermostat.
There is in consequence no center of hydrodynamics at which the hydrodynamic friction would
apply and, in conjunction with the opposite self-propelled force, cause a torque on the
colloid.
The trajectories in Fig.~\ref{l-sp-six-noh} do not show any sign of circling behaviour,
highlighting that, even for single-particle studies, hydrodynamically resolved fluids play
an important role in simulation models.

\section{Conclusions}
\label{conclusions}

Chemically powered nanomotors represent promising devices to autonomously execute tasks at
the nano- and
micro-scale~\cite{katuri_swimmer_applications_2017,ebbens_opinion_2016,wang_nanomachines_2013,de_avila_immunoassay_2017,baraban_cargo_2012,aubret_opinion_2017}.
Among the many challenges to resolve is the control of individual nanomotor trajectories as
a function of their shape and functionalization.
It is thus beneficial for the chemical, physical and engineering communities to improve the
abilities of our computational modeling toolkit.
Mesoscopic simulations have demonstrated their usefulness for understanding the mechanisms
of
chemical~\cite{yang_wisocky_ripoll_2014,valadares_el_al_sphere_dimers_small_2010,ruckner_kapral_prl_2007,reigh_kapral_soft_matter_2015,kapral_perspective_jcp_2013,de_buyl_kapral_nanoscale_2013}
and thermal self-propulsion~\cite{de_buyl_kapral_nanoscale_2013,yang_ripoll_pre_2011}.

Building on the work of Rückner and Kapral~\cite{ruckner_kapral_prl_2007} and on rigid-body
Molecular Dynamics~\cite{rozmanov_rotational_2010}, we extended the applicability of
coarse-grained nanomotor simulations to the L particle.
Considering first passive Brownian motion, we showed the coupling between the degrees of
freedom, when tracking the center of mass of the colloid. By locating the center of
hydrodynamics~\cite{chakrabarty_brownian2d_2014}, we were able to decouple the rotational
degrees of freedom from the translational ones and to obtain the actual lever length for the
self-propulsion induced torque.
In the active motion of the L particle, we find the same circling type of trajectories as in
Ref.~\cite{kummel_circular_prl_2013}.

In the present work, the mobility matrix of the particles emerges from the colloid-fluid
interaction, so that the circling patterns results directly from the shape of the particle
without further input to the simulation procedure.
To our knowledge, those are the first such simulation results for the symmetry class
``asymmetric boomerang'', following the naming scheme of Chakrabarty {\em et
  al}~\cite{chakrabarty_brownian2d_2014}.
As such, our results directly support the use anisotropic colloids for the study of
(i) many-motor systems~\cite{colberg_kapral_manybody_2017}, (ii) the inclusion of external
fields (flows and/or chemical gradients)~\cite{deprez_chemotactic_2017}, or (iii)
sedimentation~\cite{ten_hagen_gravitaxis_l_2014,kuhr_sedimentation_2017},
either in support or in preparation of experiments.

\section*{Acknowledgements}

Pierre de Buyl is a postdoctoral fellow of the Research Foundation-Flanders (FWO).
The author wishes to thank Enrico Carlon, Peter Colberg, Thibaut Demaerel, Carlos
Echeverria, Pavlik Lettinga, Stefanos Nomidids, and Snigdha Thakur for many discussions, and
Christian Maes for support.
The author is grateful to Raymond Kapral for initiating this line of research, for many
interesting discussions, and for comments on the manuscript.

\appendix

\section{Computational reproducibility}
\label{repro}

We describe here the practical computational details to reproduce the present work.

We ran all the simulations with the program \texttt{single\_body} from the open-source
package RMPCDMD for the simulation of passive and chemically active
colloids~\cite{de_buyl_rmpcdmd_2017}. RMPCDMD is implemented in Fortran 2008 and uses OpenMP
for multithreaded operation.
RMPCDMD outputs the trajectories in H5MD~\cite{h5md_cpc_2014} files, a HDF5-based
specification for molecular simulation data.

For the analysis, we used the Python programming language with the following libraries:
NumPy~\cite{numpy_csie_2011} for the basic numerical layer, SciPy~\cite{scipy-web} for
numerical integration, matplotlib~\cite{matplotlib_2007} and Mayavi~\cite{mayavi_2011} for
the figures, h5py~\cite{collette_python_hdf5_2013} to read HDF5 files.
We computed the mean-square displacements and cross-displacements with the package
tidynamics~\cite{tidynamics_2018}.
We wrote the analysis of the simulation data in Jupyter
notebooks\footnote{\url{http://jupyter.org/}} to the level that all the figures can be
reproduced.

We uploaded the parameter and coordinate files and the notebooks for the analysis and
figures to the Zenodo archival platform~\cite{complex_nanomotors_2018}.

\bibliographystyle{unsrtnat}
\bibliography{/home/pierre/code/bibfile/pdebuyl}

\end{document}